# GRETA: Modular Platform to Create Adaptive Socially Interactive Agents

Michele Grimaldi
ISIR, Sorbonne University
Paris, France
michele.grimaldi@isir.upmc.fr

Jieyeon Woo
ISIR, Sorbonne University
Paris, France
jieyeon.woo@isir.upmc.fr

Fabien Boucaud
ISIR, Sorbonne University
Paris, France
fabien.boucaud@isir.upmc.fr

Lucie Galland
ISIR, Sorbonne University
Paris, France
lucie.galland@isir.upmc.fr

Nezih Younsi
ISIR, Sorbonne University
Paris, France
nezih.younsi@isir.upmc.fr

Liu Yang
ISIR, Sorbonne University
Paris, France
liu.yang@isir.upmc.fr

Mireille Fares
ISIR, Sorbonne University
Paris, France
mireille.fares@isir.upmc.fr

Sean Graux
ISIR, Sorbonne University
Paris, France
sean.graux@isir.upmc.fr

Philippe Gauthier
ISIR, Sorbonne University
Paris, France
philippe.gauthier@isir.upmc.fr

Catherine Pelachaud
CNRS, ISIR, Sorbonne University
Paris, France
catherine.pelachaud@upmc.fr

## ABSTRACT

The interaction between humans is very complex to describe since it is composed of different elements from different modalities such as speech, gaze, and gestures influenced by social attitudes and emotions. Furthermore, the interaction can be affected by some features which refer to the interlocutor's state. Actual Socially Interactive Agents SIAs aim to adapt themselves to the state of the interaction partner. In this paper, we discuss this adaptation by describing the architecture of the GRETA platform which considers external features while interacting with humans and/or another ECA and process the dialogue incrementally. We illustrate the new architecture of GRETA which deals with the external features, the adaptation, and the incremental approach for the dialogue processing.

## 1 INTRODUCTION

During an interaction, human partners display a wide range of behaviours that are tight to the communication process. Partners' facial expressions, hand gestures, gaze patterns, etc., participate in conveying intentions and affects, managing speaking turn exchanges, and building rapport, to name a few functions.

Socially interactive agents (SIAs) are virtual or physical entities designed to interact verbally and non-verbally with their human interlocutors [38]. They are endowed with a body, be cartoonish or realistic. They are modelled to perceive their interlocutors, reason, and decide what to say and how; they can express a wide variety of affects and social attitudes. They are entering more and more our everyday lives, being used as information providers [39], tutors [8], medical assistants [9], or even as virtual friends [12].

Building a system that can act as an SIA requires much multidisciplinary effort. It involves natural language understanding and processing, dialogue manager, emotion modelling, animation, speech synthesis, nonverbal behaviour modelling, etc. To overcome the complexity of building an SIA platform, the SAIBA architecture has been proposed [31]. It consists of three main components, each dedicated to a specific task: the Intention Planner plans the communicative intentions and affects to be conveyed by the SIA, the Behavior Planner instantiates these functions into multimodal behaviours, and the Behavior Realizer computes the corresponding animation. Two representation languages have been defined to pass messages between these modules. The Function Markup Language FML encodes the communicative intentions and emotions [29] and the Behavior Markup Language BML captures the multimodal behaviors [56].

We have developed an SIA platform, the GRETA platform, that complies with the SAIBA framework. GRETA is modular allowing combining modules to implement different use-cases. SIAs can be driven at several levels such as the communicative and emotional functions and the multimodal behaviors they should display. Lately, we have been working on adding reactivity and dynamism to the SIA capacities. This involves considering a wide range of characteristics involved in human interaction, such as speech, gaze, expression, emotions, etc. For this reason, we have added:

- a component that allows the SIA to be sensitive to external features such as the human gaze and the virtual social touch
- a reciprocal adaptation loop to adapt the agent's behaviour according to the behaviour of its human interlocutor

In addition, we changed the way the signals to be displayed by the SIA are scheduled, introducing an incremental approach.

In the next sections, after the state-of-the-art section, we present the overall architecture of the GRETA platform. The following sections are devoted to the description of its different modules. We end the paper by describing several use cases based on the GRETA platform.

## 2 RELATED WORKS

An "*agent platform*" is a platform that systematically integrates a set of capabilities to cover various essential features of a Socially Interactive Agent (SIA) while allowing for future expansion [27]. One category of agent platforms comprises cognitive architectures that aim to simulate different aspects of the human mind into an agent's mind model. One of the earliest cognitive architectures is *Soar* [33], an architecture that uses symbolic reasoning, goals, and operator rules to generate behavior, incorporating different memory types and learning mechanisms. *Soar* has been applied successfully in the development of intelligent systems with compatibility for multiple programming languages through the Soar Markup Language (SML) and a suite of development tools. Later on, *ACT-R* [1] was proposed as a Soar-inspired cognitive architecture utilizing symbolic reasoning, and applied in educational applications like intelligent tutoring systems. *OpenCog* [25] was later on proposed as a platform for general intelligence, combining cognitive algorithms referred to as "cognitive synergy". *TinyCog* [5] was also proposed as a simplified version of a cognitive framework. It follows a concept called Scene Based Reasoning, where scenes simulate real-world settings. Recently, *Sigma* [43] was developed to achieve grand unification, generic cognition, functional elegance, and sufficient efficiency.

Another category of platforms was developed to cover a wide range of specific capabilities related to SIAs. The Virtual Human Toolkit (VHToolkit)[26, 28] is a comprehensive collection of approaches and technologies that have been used in a great number of projects such as MRE [42], SASO [52], and SGT Star [2], integrating their architecture, nonverbal behavior technology, Natural Language Processing (NLP), and rendering technologies into a unified platform called the Gunslinger project [28]. VHToolkit follows the SAIBA framework and employs a modular architecture. Modules communicate with each other using a custom protocol called VHMsg, built on top of ActiveMQ. They can be implemented in languages like C#, Java, and C++, allowing for the incorporation of new modules as long as they adhere to the VHMsg protocol. Although the VHToolkit primarily supports Windows, there is ongoing development for a multiplatform version. It includes PocketSphinx as the default speech recognition solution, with options for Google ASR and native Windows 10. Audio-visual sensing is provided by MultiSense, which utilizes the SSI framework. NLP is facilitated by the NPCEditor, a statistical text classifier that matches user input to generate appropriate character responses. The NPCEditor can also incorporate custom Groovy scripts for dialogue management. Nonverbal behaviors are generated by the NonVerbal Behavior Generator (NVBG), which receives FML from the NPCEditor and creates a BML schedule based on syntactic and semantic rules. It includes SmartBody, a powerful procedural character animation and simulation platform responsible for realizing the BML schedule. Text-to-speech functionality defaults to Festival but also offers options such as CereVoice and MS SAPI. The rendering capabilities of the VHToolkit are supported by Unity.

Other platforms have been developed. The Relational Agents platform [6] consists of 2D virtual agents commonly utilized in the healthcare sector. The framework includes a task planner, a dialogue manager with associated ontology, a web-based Behavior Markup Language (BML) realizer, and text-to-speech integration. It is implemented using Java and Flash. The platform WASABI [4] was developed to simulate both primary emotions resembling infants and secondary emotions with cognitive elaboration by combining physical emotion dynamics with cognitive appraisal. The Virtual People Factory [44] platform was proposed as a web authoring and runtime platform for creating virtual patients. Visual SceneMaker [22] is another platform used as an authoring tool, it enables non-experts to develop interactive presentations and has been utilized in various projects. ADAPT [49] platform focuses on full-body animation, navigation, and object interaction by integrating various approaches and technologies, including SmartBody, into a unified framework. The Articulated Social Agents Platform (ASAP) [61] provides a collection of software modules for both SIAs and Social Robots (SRs). ASAP complies with the SAIBA framework and includes Flipper2.0 [54][55] which provides a technically stable and robust dialogue management system to integrate with other components of ECAs such as behaviour realisers. The primary programming language used is Java. The Generalized Intelligent Framework for Tutoring (GIFT) [50] is another platform that primarily focuses on delivering intelligent tutoring capabilities for both desktop and web platforms. It incorporates some SIA capabilities through integration with a subset of the VHToolkit. M-PATH [62] emphasizes empathetic conversations and combines SmartBody with a custom dialogue manager. The Standard Patient Studio [53] was proposed to enable doctors and medical students to create and practice with their own standardized patients. It is an online authoring tool that begins with an interactive, healthy patient as a baseline and allows for deviations. It also includes student feedback capabilities.

On the other hand, *game engines* platforms were built to design animated characters as well as their behavior. They have different functionalities such as animation, sound, networking, rendering, and much more. They are designed to be compatible with multiple hardware platforms like mobile, web, AR, and VR, and provide adaptable development environments. The two most widely used game engines are Unity and Unreal Engine. Unity initially targeted developers that are beginners while Unreal Engine targeted large professional teams and created highly realistic graphics. They both aim to benefit researchers and developers to create SIAs.

## 3 THE GRETA PLATFORM

GRETA [41] is a real-time multi-platform aimed at controlling the behaviours of socially interactive agents (SIAs) with 3D models compliant with the MPEG-4 animation standard [40]. More specifically, the GRETA platform specializes in the multimodal animation of humanoid characters: it can produce simultaneous facial expressions and body gestures and combine them with head (for adjusting gaze direction) and torso movements. Speech can also be produced via text-to-speech modules. GRETA is based on the standard SAIBA architecture [31] and allows for a semantic approach to the control of SIAs via the use of the two standard XML formats of FML [29] and BML [56] which are aimed at the encoding of, respectively, the communicative intentions and the multimodal behaviours expected from the agent. From the software point of view, the core of GRETA's design philosophy is the notion of modularity. In the GRETA application, almost any component can be removed and replaced by another similar but more specialized component. This allows developers to build different modules to tackle different research goals, as long as those modules can conform to the overall SAIBA process. In the next sections, we present the core architecture of the GRETA platform followed by its various components.

## 3.1 Core architecture

At its core, GRETA is a platform aimed at simulating social behaviours via the control of the animations of 3D character models. Core components of the platform therefore include the multimodal signals creation tools. Animation can be crafted in the application thanks to graphical modules which allow the user to manipulate the 3D model of the agent and record the sequence of signals (for example keyframes or action units [13], depending on the type of animation we want to create) we build. Those resulting gestures, facial expressions, torso movements, etc. which we call behaviours, can then be named and recorded in specific library files such as the gestuary (where we will store all the definitions of arm gestures) or the facial expressions library. This allows us to ultimately create a lexicon, a library where we pair different sets of behaviours with a communicative intention they can represent. Think for example that the *anger* emotion could be expressed via the combination of a facial expression featuring a frown and some ample arm gestures. FML files can then reference those pre-defined communicative intentions (*anger*) and be processed to produce the corresponding BML files (behaviour level, the frown, and the ample arm gestures) which can also be processed to produce the actual signals to be rendered (action units for the face and keyframes for the body, corresponding to the gestures). As a result, a minimal example of the most standard way of using GRETA would be first preparing an FML file, in which we will write an utterance the agent should say and specify for each part of the utterance which intention is being expressed at this particular moment in time. Next, this FML file is sent to GRETA, which processes it and computes the corresponding BML tags that specify which behaviors (gestures, facial expressions, etc.) to display, to convey each intention, and synchronize them with speech. Finally, this BML file is processed again to produce the actual signals that correspond to the production of the speech and the animations, which can then be rendered via a voice synthesizer and a 3D engine. This process is the core of the GRETA platform and its most basic functionality. It allows simulating virtually any human communicative or expressive behaviour, but it is not yet enough to be able to qualify GRETA-controlled agents as socially interactive agents. We still need to enable interactivity and adaptation. The following sections illustrate how the basic GRETA architecture has been enhanced over the years to allow GRETA-controlled agents to perceive human users' behaviours towards them and automatically adapt their behaviours accordingly.

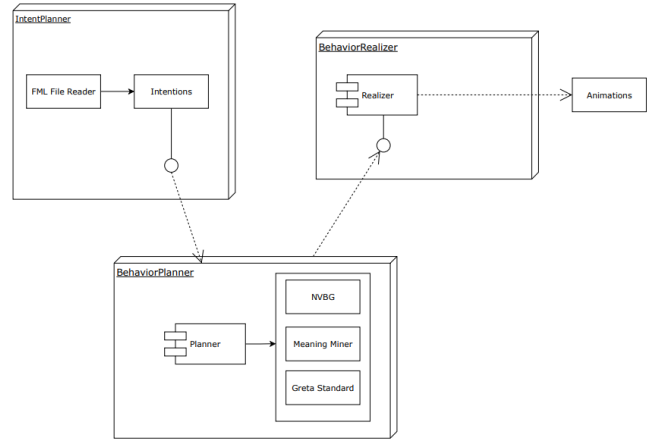

Figure 1: GRETA's Architecture Schema w.r.t. [24]

## 3.2 User's multimodal signals

The GRETA architecture can take several *input signals* to capture information related to human users interacting with GRETA virtual agents. These raw signals can be users' facial expressions, head movements, speech but also hand landmarks and proximity. These signals are captured through a variety of media including a camera, microphone, Leap motion device to capture hand landmarks, and a virtual reality VR tracking system to obtain proximity values.

At the *signal processing* level, we rely on existing open-source modules.

*Facial expression and body movement.* Visual features of face Action Units (AUs) [13] and facial landmarks, head movements, and gaze are extracted from the user face (i.e. user's facial expressions and head movements) via OpenFace [3]. OpenFace is an open-source toolkit for facial feature extraction while OpenPose [] can extract body joints. Mediapipe [37] which is also an open-source framework for facial feature extraction can also be employed.

*Speech.* The user's utterance transcription is obtained from the user's speech signal using Google ASR [1]. ASR (acronym for Automatic Speech Recognition also known as STT: Speech-to-Text) is the technique of transcribing spoken words into text. ASR identifies full phrases and converts them into text during the user's speech. For the prosody, it is rendered by openSMILE [14], an open-source toolkit for audio feature extraction. Various prosodic features such as the fundamental frequency, loudness, voicing probability, and Mel-frequency Cepstral Coefficient (MFCC) [36] can be obtained in real-time.

## 3.3 User's signals interpretation

All these modules output *signals* (e.g., Actions Units [13], acoustic features) that are sent to the *signal interpretation* modules. For instance, from the text obtained by a speech recognizer we can extract image schemas [], dialog acts [], or automatic thought []. These signals are then used to compute the multimodal behaviours of SIAs. We detail now how these specific extractions are obtained.

---
[1]https://cloud.google.com/speech-to-text

**BERTIS.** *BERTIS* is an Image Schema computational model designed to classify an input text into an Image Schema class. Image Schemas are cognitive structures that represent recurring patterns in sensory-motor experiences. They play a significant role in human cognition and language understanding. To compute the Image Schema class for a given input text, *BERTIS* employs the pre-trained BERT model [10]. More specifically, the BERT model is fine-tuned specifically for the task of classifying the input text into an Image Schema class. The training data for *BERTIS* is obtained from the Image Schema Corpus introduced by Wachowiak et al. [57]. The corpus contains a diverse set of text examples paired with their corresponding Image Schema classes. This model is currently found at https://huggingface.co/mireillfares/BERTIS. *BERTIS*'s ability to classify input text into relevant Image Schema classes is used in GRETA's gesture generation process, allowing GRETA to incorporate the appropriate Image Schemas when generating gestures from the given input text (see Section 5).

**Dialog acts** A dialog act labels an utterance according to its function in a dialog, they reflect the act the speaker is performing. Dialog acts can be extracted through automatic classifiers such as dialogtag. Dialog tag is based on hugging's face DistillBert base [46] fine-tuned on a subset of the Switchboard-1 corpus [23] consisting of 1155 conversations. The resulting model classifies utterances between 42 tags including statement-non-opinion, acknowledge, statement-opinion, agree/accept, etc.

*GPT3.* GRETA has been enhanced with the integration of a new module designed specifically to handle OpenAI's GPT-3 technology [7]. The module consists of two main components: a Java-based server and a Python client. The Java server is responsible for receiving questions from users submitted through the GRETA graphical user interface (GUI), processing the questions, and sending them to the Python client. The Python client then calls the GPT-3 API to generate an answer and then sends it back to the Java server. The answer is finally formatted in an FML file which is used to control its nonverbal behavior and facial expressions while answering the question. This allows for a more natural and engaging interaction between the user and the virtual agent.

## 4 INTENT PLANNER

The Intent Planner is designed to render the agent's communicative intentions and emotions. Its role is to compute motives and objectives guiding the agent's behaviour during the whole interaction. To do this computation the planner can take in input different signals from the user such as speech, gaze, and gestures.

### 4.1 FAtiMA

To determine the intention the agent should adopt in response to a user's inputs, we can combine the GRETA platform with an external computational decision model that would act as an Intent Planner. We use the FAtiMA (Fearnot AffecTIve Mind Architecture) Toolkit, a computational model of emotions architecture tailored to generate an emotional state for the agent. This emotional state can then modulate the agent's intentions and behaviour. FAtiMA is based on the cognitive appraisal paradigm and uses inference rules based on logic programming. Those rules are applied based on the agent's knowledge of the interactive situation, which is dynamically stored in a knowledge base when new users' actions and movements are detected. The decision rules are designed based on the literature on human-human social touch to determine when the touch modality would be both coherent and acceptable to the user. Specifically, we designed interpretation rules to evaluate the level of rapport between the human user and the agent, so that the decision model may first determine which communicative intention to express (coherence) and then take the level of rapport into account to decide whether to use touch or not for this specific intention (acceptability). The final decision can then be passed onto GRETA via an FML or BML file, where the exact gestures (including the touch gestures if there are any) to use for the specified intentions and modalities will be selected.

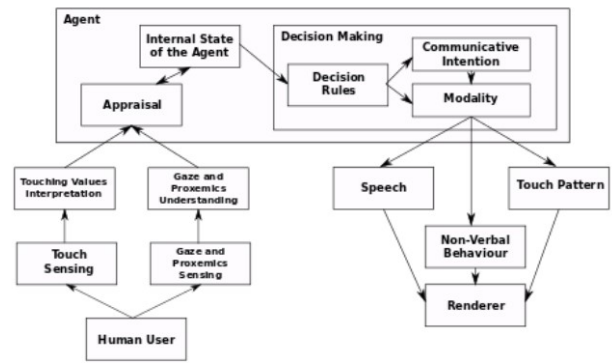

**Figure 2: Computational model of emotions**

### 4.2 Adaptive dialogue planner

The GRETA framework can be utilized in conjunction with more intricate dialogue planners, as demonstrated in [21]. To simulate adaptation mechanisms that happen during interaction, we have proposed an adaptive dialogue planner. This dialogue planner can employ either task-related or social-related dialogue acts and select the next dialogue act to maximize user engagement. User engagement is detected using verbal and nonverbal behaviour cues. The adaptation process is performed through reinforcement learning, with an adaptation module implemented in Python outside of GRETA. ActiveMQ is used to transmit the user's multimodal input to the Python module. Subsequently, the Python module selects the optimal dialogue act and nonverbal behaviour in terms of warmth and competence, which is then relayed back to GRETA via ActiveMQ. The basic nonverbal behaviour of the agent is generated using Meaning Miner, as discussed in Section 5.

## 5 BEHAVIOR PLANNER

The Behavior Planner in the GRETA system encompasses two key modules, namely Meaning Miner [24] and NVBG (Nonverbal Behavior Generator) [34], each contributing to the synthesis of nonverbal communication. The Meaning Miner module, rooted in Image

Schemas and Ideational Units, acts as a link between verbal and nonverbal channels. Processing FML files enriched with prosodic and Ideational Unit markings, it identifies relevant Image Schemas, aligns them with speech, and generates corresponding gestures. This involves extracting gesture invariants, determining features for each Image Schema, harmonizing gestures within Ideational Units, and dynamically adjusting gesture properties for expressive coherence. On the other hand, NVBG is a component developed by Marsella and his team. NVBG extracts semantic functions (e.g. affirmation, negation) from a text. Based on a corpus analysis, a set of mapping rules was designed to map these features into multi-modal behaviours. NVBG is connected to GRETA. It augments its actions by comprehensively analyzing the agent's text, its semantics, and emotional context to generate nonverbal actions. Finally, the GRETA Behavior Planner translates the FML tags into BML tags and sends them to the Behaviour Realizer in charge of executing the behaviours.

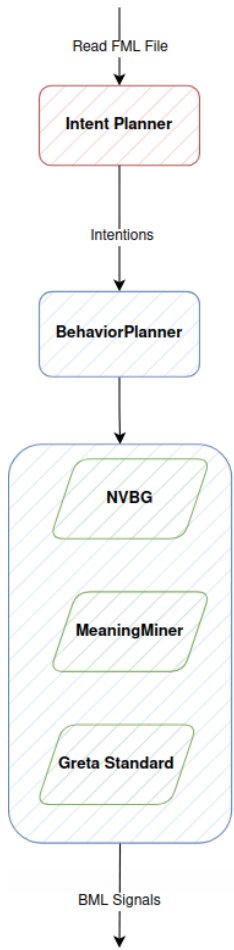

Figure 3: Behavior Planner Schema

## 6 BEHAVIOR REALIZER

The Behavior Realizer generates the animation for the agent following the MPEG-4 format. The inputs are specified by the BML language. It contains the text to be spoken and/or a set of nonverbal signals such as facial expressions, gaze, gestures, and torso movements, to be displayed. Each BML tag is transformed into a set of keyframes that are then smoothly interpolated. The Behavior Realizer ensures the synchronization of behaviours across modalities and resolves conflicts between signals using the same modality. Finally, the speech is generated by an external TTS (Text-to-Speech) system, such as MaryTTS and Cereproc, with lip movements added to the animation.

### 6.1 Incremental Behavior Realizer

Dialogue processing is, by its very nature, incremental. No dialogue agent (artificial or natural) processes whole dialogues, if only for the simple reason that dialogues are created incrementally, by participants taking speaking turns. At this level, most current implemented dialogue systems are not incremental (except for [32]): they process user utterances as a whole and produce their response utterances as a whole. Incremental processing, as the term is commonly used, means more than this, however, namely that processing starts before the input is complete [30]. Incremental systems hence are those where "each processing component will be triggered into activity by a minimal amount of its characteristic input" [35]. If we assume that the characteristic input of a dialogue system is the utterance, we would expect an incremental system to work on units smaller than utterances. Doing this then brings into the reach of computational modelling a whole range of behaviours that cannot otherwise be captured, like concurrent feedback, fast turn-taking, and collaborative utterance construction.

The SAIBA framework allows the generating of multimodal behaviours for sequences of intentions, often at the speaking turn level. It lacks the flexibility to manage reactive behaviours such as those requiring fast adaptation due to an interruption, or a repair for example. Incremental processing has certain benefits when compared to non-incremental processing, which only takes into account full utterances or even longer units such as turns. Incrementality can be beneficial when attempting to optimize the reactivity of a system. For instance, if an input is being processed, beginning the process before the input is finished can result in a faster completion time than if the process was only started after the input was complete. In the past years, efforts have been made to develop incremental dialogue manager [47], speech recognition, and synthesis [32]. Kopp and colleagues have proposed an architecture which combines rule-based, behavioural feedback responses to speaker elicitation events with the notion of a "concerned", collaborative listener that strives to keep track of what a speaker is saying to compute incrementally multimodal feedbacks.

We have modified the GRETA architecture to ensure the incremental realization of nonverbal behaviours. The incrementality module relies on the concept of chunks (similar to the concept of Incremental Units introduced in [48]). These are groups of keyframes. They have a time period and contain all the keyframes within that time period. The Incrementality module is composed of the following modules:

- **Incremental Realizer**: It is a modified version of the Behavior Realizer that takes in input, a list of signals, generates keyframes, and separates them into chunks (groups based on keyframe starting time). It then uses a dedicated thread to send keyframe chunks one by one according to time unit.
- **Incremental Realizer Interaction**: This module interacts with the Incremental Realizer by sending interruption commands.

*6.1.1 Incremental Realizer.* The Incremental Realizer transforms a list of signals sent by the Behavior Planner into a list of keyframes. It then forms a list of keyframes chunks based on their starting time and sends them to a thread scheduler which schedules these chunks based on their starting time, waiting between chunks to ensure proper timing.

## 6.2 Incremental Realizer Interaction

The Incremental Realizer Interaction module is used to interact with the Incremental Realizer module through the following actions.

- **Interrupt**: It halts the transmission of subsequent chunks.
- **Resume**: Upon receiving an Interrupt command, the sending chunk action will be resumed from the last one before the interruption.
- **Stop**: The execution is completely halted and the voice is silenced, returning to the resting position.
- **Clear Thread Queue**: Empty the chunk list and close the queue in the thread.

## 6.3 Frame-level Behavior Realizer

The Frame-level Behavior Realizer is a critical module designed to enable the virtual agent to exhibit smooth and responsive behaviour in real-time. Its primary purpose is to process the agent's behaviour frame by frame, ensuring that the interaction remains fluid and continuous. By leveraging this module, the virtual agent can adapt dynamically to user input and computational model outputs, providing a more natural and engaging interaction experience.

At each time step, the module receives behaviour signals generated by an upstream computational model (e.g., ASAP). These signals are communicated via the Open Sound Control (OSC) protocol. Unlike the Behavior Realizer, which focuses on planning and executing agent behaviours in pre-defined chunks, the Frame-level Behavior Realizer operates at a finer granularity, directly utilizing raw user input data to generate real-time behavioural adjustments.

The types of behaviors that the virtual agent can display include:

- Upper face Action Units (e.g., *AU1*, *AU2*, *AU4*, *AU5*, *AU6*, *AU7*);
- Smiles (*AU12*);
- Blinks (*AU45*), which are automatically generated;
- Head movements along the x, y, and z axes;
- Gaze movements along the x and y axes;
- Mouth movements.

To configure the parameters and select the behaviours to be displayed, an interactive interface is provided. This interface allows users to specify the initial settings for the module, including the activation or deactivation of specific behaviours. It is important to note that the interface serves only as a configuration tool and is distinct from the core functionality of the Frame-level Behavior Realizer, which operates seamlessly in real-time during the interaction

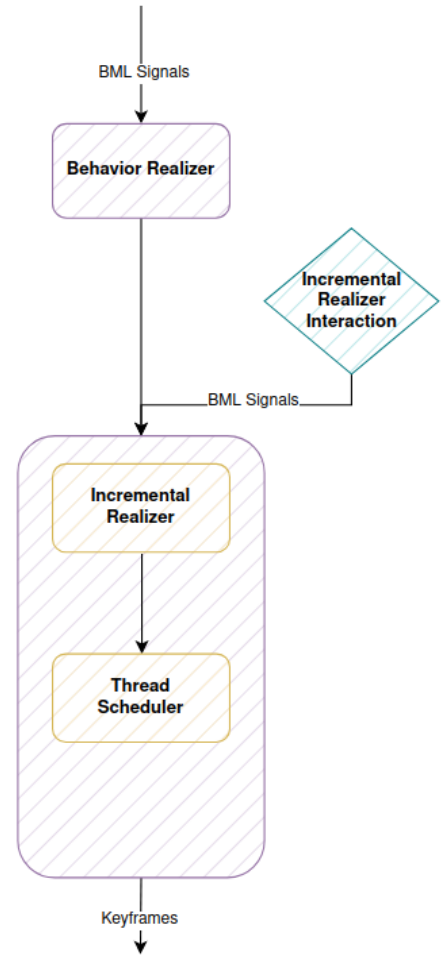

Figure 4: Incremental Realizer and Interaction Schema

## 7 USE CASES

### 7.1 Behavior Generation and Adaptation

The GRETA platform, as integrated with the Flipper 2.0 dialogue manager 6, supports the development of a virtual motivational interviewing (MI) counsellor. This use case leverages advanced behaviour generation capabilities to create an agent that engages users through both verbal and nonverbal interactions, following motivational interviewing and cognitive-behavioural therapy (CBT) principles. The system architecture, illustrated in Fig. 6, is designed to process multimodal input from the user and generate context-adaptive behaviour outputs in real-time.

*System Architecture and Input Processing.* To capture the user's input, the system incorporates:

- **OpenFace module**: Extracts facial Action Units (AUs) to analyze the user's nonverbal cues.

- **ASR module (Automatic Speech Recognition)**: Converts the user's speech into text in real-time.

The input data, including action unit activations and textual content, is transmitted to Flipper 2.0 using socket-based communication. The Flipper 2.0 module processes these multimodal signals and generates the agent's behaviours (in FML format), which are rendered by the GRETA platform.

*Behavior Selection and Generation.* The agent's behaviour generation mechanism builds upon motivational interviewing principles to produce context-sensitive and engaging responses. Previously, a **rule-based approach** was employed, where frequent patterns of co-occurrence between user and counsellor behaviours were identified through a detailed analysis of an MI corpus. This approach allowed the system to map user input (both verbal and nonverbal) to corresponding agent behaviours based on predefined patterns, ensuring basic context adaptation and adherence to MI principles.

However, to overcome the limitations of rule-based systems and enhance adaptability, the system architecture is transitioning toward a **data-driven approach**:

- **Reinforcement Learning (RL) Integration**: A reinforcement learning model is being developed to communicate directly with Flipper 2.0. This model will predict and select the most contextually appropriate behaviours for the agent based on reward estimations derived from multiple factors, including:
  - Action Unit activations,
  - Textual indicators of social rapport,
  - Contextual data from the ongoing interaction.
- **Behavior Adaptation**: The RL model enables the agent to refine its behaviour selection over time, learning from interactions and dynamically adapting to the user's emotional and conversational state.

*Integration of Motivational and CBT Principles.* To ensure the agent aligns with therapeutic goals, the system integrates motivational interviewing (MI) and cognitive-behavioural therapy (CBT) frameworks. These principles guide both the behaviour generation and the adaptation processes, ensuring that the agent's responses are not only contextually relevant but also conducive to fostering engagement and building rapport.

By reorienting behaviour generation around a hybrid rule-based and data-driven architecture, the GRETA platform can better address the complexity of real-time interactions while offering a scalable foundation for future advancements in virtual agents.

### 7.2 Real-time Adaptive Behavior Generation

In a conversation, it is important to show nonverbal behaviors which can be accompanied by verbal messages or can be used alone to communicate intentions. While communicating with our interlocutors, we not only send social signals to convey our message but also adapt to their signals. The integration of this reciprocal adaptation, allowing social and engaging conversations, to SIAs can permit them to interact with human users in the same way. The adaptation should be done continuously for better engagement of the users. We extended the GRETA platform and proposed the Interactive and Adaptive Virtual Agent (IAVA) system [60] which allows the real-time computation of the agent behaviour from the behaviour of its human interlocutor and simulation of the dynamic behaviour adaptation between them. It focuses on two aspects which are real-time adaptive behavior generation and natural dialogue management

The real-time adaptive behaviour of the agent is generated via the Augmented Self-Attention Pruning (ASAP) model [61] which endows the agent with reciprocal adaptation capability and renders the next agent's behaviour at the frame-level of $25 fps$ using the previous visual and audio features of both the human user and the agent. The user and agent's features (speech prosody and facial gestures) are perceived through OpenFace and openSMILE. The ASAP model receives this information via the communication protocols of ActiveMQ, ZeroMQ, and OSC to predict the agent's behavior at the next frame. The inference speed of the ASAP model is approximately $0.008s$. The obtained agent behavior is displayed for every frame via the Frame-level Behavior Realizer, presented in Section 6.3. The IAVA integrates the ASAP model and Frame-level Behavior Realizer within its Adaptation Behavior Realizer module.

The natural flow of the conversation between the agent and the user is also managed by the IAVA system via its Dialogue Manager module. The module controls the dialogue using Flipper2.0 which chooses the communicative intentions (via agent utterance) and directs the conversational flow with its Turn-taking Management module which handles conversational turn-taking via a rule-based technique that looks at the speaking states of the agent and that of the human user.

The agent's utterance is instantiated into mouth movements and synchronized with the agent's speech via the modules of Behavior Planner, Behavior Realizer, and Speech Synthesizer. The produced agent's adaptive facial gestures and mouth movements are sent to the Ogre3D for visualization and the agent's speech is rendered with the Speech Synthesizer.

The IAVA system functions in real-time, with an execution time of $0.04s$ of a single system loop which consists of:

- perception of approx. $0.03s$ with OpenFace at $30 fps$ and openSMILE at $100Hz$;
- adaptive behaviour generation of approx. $0.008s$ via the ASAP model;
- communication and visualization of approx. $0.002s$.

The frequency of $25Hz$ (i.e. $0.04s$) is assured for the generation and display of the agent behaviour by syncing all signals within the system without any delay.

The system can be used for various applications. By switching to another scenario, a simple change of FML messages (or the agent's pre-scripted utterances), the corresponding adaptive nonverbal behaviours can be generated. The IAVA system has showcased its benefit of producing adaptive agent behaviour for Cognitive Behavioral Therapy (CBT) [59] and Social Skills Training (SST) [45].

Furthermore, the computational model can be exchanged with another model generating the agent's facial expression and head/gaze movements at the frame-level. A deeper computational model, AMII model [58], which also renders the same agent behaviour for each frame by capturing intrapersonal relationships (of the agent and human) along with the interpersonal relationship and modelling the multimodal information at a deeper level, can be employed with the IAVA system.

## 7.3 Social Touch

To further enhance the communicative abilities of the GRETA-controlled agents, we explored a currently under-represented modality of communication: social touch. Taking inspiration from the literature on human-human social touch and drawing from the field of haptics we extended the GRETA architecture with a framework allowing our agents to both touch and be touched in interaction with a human. In 6 the integration with the whole architecture is depicted. The framework is based on the following elements:

- Virtual Reality and haptic feedback:
  - Virtual reality headset allows direct 1:1 scale interactions with the agent
  - The Leap Motion camera from Ultraleap detects and reproduces the user's hand gestures
  - A vibrotactile sleeve generates haptic stimuli on the user's am
- GRETA's social-touch recognition module
- GRETA's FeedBack Module
- Unity environment
- A perception module to detect the human's actions and touch
- FAtiMA Toolkit [11] for emotional decision making

*7.3.1 Virtual Reality Implementation.* Our framework is currently implemented in a Unity application played inside a HTC Vive headset. To allow 1:1 scale interactions, especially touch interactions, we required some way to not only track the head of the user but also their hands. We achieved this by using the Ultraleap solution of the Leap Motion cameras that can track hands and interface with Unity to reproduce the user's hands and their movements in the virtual environment. By combining the head-mounted display (HMD) with the Leap Motion controllers, we built a virtual avatar for the user: the tracking of the head by the VR headset allows us to correctly place the avatar in the virtual environment, the Leap Motion allows us to co-localize their hands, and inverse kinematics-based animations allow us to combine both information to animate and co-localize a complete humanoid avatar for the user. Furthermore, the Ultraleap Unity plugin includes an interaction framework that enables manual interactions with some game objects (grasping, pushing, throwing,...). On the other hand, the plugin is not suited for social interactions as its current version cannot prevent the hands from going through static objects such as an agent's 3D model. To allow realistic-looking social touch interactions, we had to simulate a proper physical interaction with collider components both on the agent's 3D body and on the avatar's virtual hands.

Unity manages collisions between objects with colliders, which can be attached to objects and define their shapes for physical collisions. Colliders are invisible and usually take the form of primitive shapes (boxes, spheres, cylinders) which can be adjusted in size and combined to approximate a complex mesh, which is more computationally (CPU) efficient. Placing colliders on the hands requires constraining them to each other to stabilize the hand shapes when the real hand of the user "entered" the 3D object while the virtual hand was blocked by the colliders of the object. Concerning the haptic feedback we use a vibrotactile sleeve. This sleeve includes several independent vibrating actuators which can vibrate with different intensities. The actuators are activated synchronously via Unity when GRETA is going to touch the human user and the intensity, number, and location of the activated motors vary according to the predefined type of touch and the place touched.

*7.3.2 Gesture Recognition and Perception modules¨: Deplacer à section User's Input.* The recognition of the type of touch performed by humans was a key component that required special implementation. Additionally, the perception module was responsible for determining the type of touch the agent had to perform. This was a crucial part of the system, as the agent needed to be able to interact with its environment. The GRETA Social-touch recognition module contains Random Forest network which will process the touch parameters coming from the Ultra Leap motion virtual hands and the implementation in Unity when the human touches the agent in the virtual environment. The network can recognize 4 types of social touch: hit, tap, caress, and stroke. The Random Forest uses the velocity, pressure, duration, and parts of the body touched and that are touching to decide to which categories the social touch belongs. To train the Random Forest we collected real data coming from the interaction between humans and the virtual agent. The model performs well on both train and test data, reaching respectively an accuracy of 0.91 and 0.80. The perception module deals with the touch gesture that the agent will perform, if it has to, and if there are the condition for it. The touch can be characterized by different parameters. We base ourselves on those defined in the literature from social touch and gesture recognition [51]. We identify six parameters of touch:

- The presence of a movement
- The intensity of the movement is determined by the kinematic velocity at the point of initial contact
- The body position is the area of the body that is being touched; the movement of the touch can be either static or dynamic, depending on whether it is moving along the agent's body
- The speed of the movement is the speed of the human's hand on the agent's body
- The duration of the movement
- The pressure the human's hand exerts on the agent's arm

Currently, we focus on only the first four parameters for our perception of touch. We do not take into account the duration of a touch gesture. We are also not able to compute the pressure of a touch.

Other used nonverbal behaviors are the proxemics which give information about the distance between the agent and the human, and the gaze direction which indicates where the human is looking. The latter parameter is also used to calculate the degree of attention of the human to the interaction. The direction of gaze is discriminated: is the human looking at the agent, at an object in the environment that has a relationship with the agent, at an object in the environment that has a relationship with the subject of the interaction, or at something else? The choices made by the agent are a direct result of the actions of the human, which have an impact on the evolution of the interactive scenario. These decisions shape the environment and the outcome of the situation. Ultimately, the human's actions are the driving force behind the development of the interactive scenario. In our experiment, there is the risk that the human is far from the agent which has a touch gesture scheduled. To deal with this event the OSC module communicates with GRETA sending the proxemics which can be classified as intimate, personal, social, or public. Then the Feedbacks module will add/remove the touch gesture computed by the FAtiMA Toolkit depending on the distance and the type of gesture.

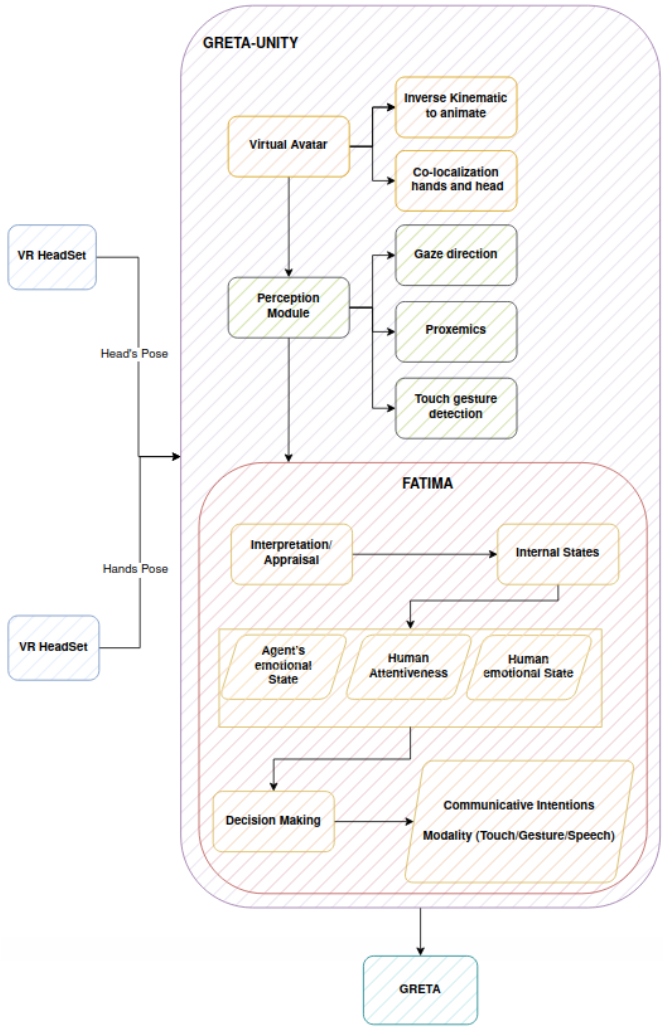

Figure 5: Social Touch Tools Schema

## 7.4 Gesture Generation Models

*7.4.1 Facial expression and head movement generation.* We propose diverse generative models for synthesizing facial gestures driven by speech and semantically-aware. We initially introduced a rule-based approach using NVBG and Meaning Miner modules [24], then a learning-based facial gestures generation model that synthesizes facial gestures based on speech prosody (the fundamental frequency F0) and text semantics (Bert embeddings)[15]. The model architecture employs an end-to-end sequence-to-sequence neural network and consists of an encoder to encode the input features and a decoder for generating the sequences of eyebrows, eyelids, and head movements. We utilized Long Short-Term Memory (LSTM) architectures to capture temporal dependencies. However, this model faced limitations due to the challenges in learning dependencies between distant positions when processing sequences. To overcome these issues, we proposed another approach for upper-facial and head gestures generation, utilizing a multimodal Transformer network [16]. This transformer network handles the sequence-to-sequence modelling of multimodal upper facial and head movements at the word level, emphasizing the correlation between head motion and upper-facial gestures to achieve a more coherent and natural agent behaviour. The resulting facial expressions and head movements are then sent to the agent which will play them as depicted in fig. 6 The proposed architecture includes a transformer network operating on multi-modal input text and speech information and a cross-attention module for efficiently exploiting semantic and speech information.

*7.4.2 Body gesture generation.* We also address the challenge of transferring the expressive style of a virtual agent to another while preserving the shape of behaviours carrying communicative meaning. Our initial solution, ZS-MSTM [18, 19], is a multimodal transformer-based approach synthesizing the behaviours of a source speaker with the style of a target speaker. We assume that behaviour expressivity style is encoded across various modalities, including text, speech, and body gestures. Our approach employs a style-content disentanglement schema to ensure that the transferred style does not interfere with the meaning conveyed by the source's behaviors. This approach eliminates the need for style labels, allowing generalization to styles not seen during training. Leveraging ZS-MSTM, we introduced TranSTYLer [17], the first multimodal generative approach synthesizing jointly facial and upper-body gestures in the style of various speakers while generalizing to those not seen during training and ensuring that the transferred style does not interfere with the meaning conveyed by the source gestures. Additionally, we proposed META4 [20], a deep learning approach generating metaphoric gestures from both speech and Image Schemas. META4 aims to compute Image Schemas with BERTIS from input text, capturing underlying semantic and metaphorical meaning, and generating metaphoric gestures driven by speech and the computed image schemas. As for the facial expressions and head movements, the resulting gestures are then sent to the agent which will play them as depicted in fig. 6 Our approach stands as the first method for generating speech-driven metaphoric gestures while leveraging the potential of Image Schemas.

## 8 CONCLUSION

In conclusion, the GRETA platform has undergone significant updates aimed at incorporating dynamic aspects of human interactions, such as speech, gaze, expression, and emotions, among others. These complex features are challenging to capture, but with advances in technology, the base architecture of the platform can be improved. The new updates to the platform include a component that allows GRETA to be sensitive to external features such as human gaze and virtual social touch, as well as the addition of a reciprocal adaptation loop. Additionally, the way different signals are scheduled has been changed, introducing an incremental approach. This approach brings a whole range of behaviors into the reach of computational modeling that would otherwise be difficult to capture, such as concurrent feedback, fast turn-taking, and collaborative utterance construction.

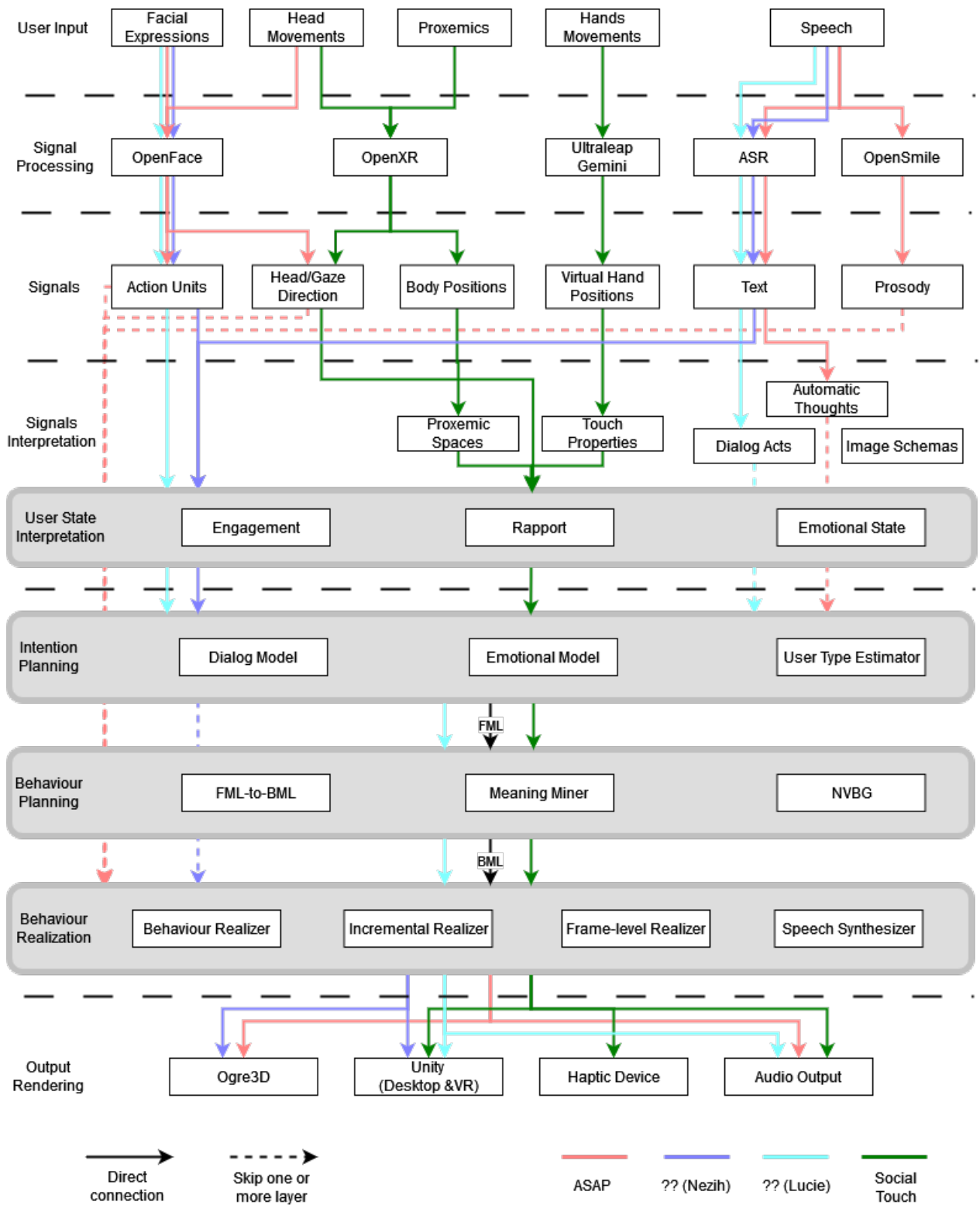

**Figure 6: GRETA's General Architecture**